\documentclass{ws-mpla}
\usepackage[super]{cite}
\usepackage{graphicx}
\usepackage{amsmath}
\usepackage{amssymb}
\usepackage{graphicx}
\usepackage{color}
\usepackage{nicefrac}
\usepackage{subfigure}
\usepackage{hyperref}

\newcommand{\note}[1]{\emph{\textcolor{red}{}}}

\newcommand{\Msun}{{\ensuremath{{M}_{\odot}}}}

\newcommand{\Cplusplus}{{\rmfamily C\raise.22ex\hbox{\small ++} }}

\newcommand{\lSect}[1]{\label{#1}}

\newcommand{\araa}{ARA\&A}%
\newcommand{\apj}{ApJ}%
\newcommand{\apjl}{{ApJ}}%
\newcommand{\aap}{{A\&A}}%
\newcommand{\mnras}{{MNRAS}}%
\newcommand{\prl}{{Phys.~Rev.~Lett.}}%
\newcommand{\nat}{{Nature}}%
\newcommand{\be}{\begin{equation}}
\newcommand{\ee}{\end{equation}}
\newcommand{\bea}{\begin{eqnarray}}
\newcommand{\eea}{\end{eqnarray}}
\newcommand{\bc}{\begin{center}}
\newcommand{\ec}{\end{center}}

\begin{document}

\markboth{Chen}
{Supermassive Black Holes}

\catchline{}{}{}{}{}

\title{Origins of Supermassive Black Holes in Galactic Centers}

\author{Ke-Jung Chen}
\address{Institute of Astronomy and Astrophysics, Academia Sinica, No.1, Sec. 4, Roosevelt Rd., Taipei 10617, Taiwan\\ Heidelberger Institut für Theoretische Studien, Schloss-Wolfsbrunnenweg 35, 69118 Heidelberg, Germany\\kjchen@asiaa.sinica.edu.tw}

\maketitle


\begin{abstract}
Direct imaging of black hole shadow halos has firmly confirmed the existence of supermassive black holes (SMBHs), with millions of solar masses, residing at the centers of the Milky Way and M87 galaxies. These groundbreaking discoveries represent a monumental success of Einstein's theory of general relativity and have revealed the hidden "monsters" lurking at the centers of galaxies. Moreover, observations of active galactic nuclei (AGNs) indicate that SMBHs with billions of solar masses were already in place within the first billion years after the Big Bang. However, the origins of these SMBHs, as well as their co-evolution with host galaxies, remain poorly understood. This review focuses on the origin of SMBHs, particularly on the formation of their seed black holes. We also highlight several outstanding challenges in modeling seed formation and discuss possible observational signatures. These signatures may be testable with current and future facilities, including the James Webb Space Telescope (JWST) and the upcoming gravitational wave observatory, the Laser Interferometer Space Antenna (LISA).


\end{abstract}

\ccode{PACS Nos.: 95.00.00.}

\section{Introduction}

Black holes were originally theoretical and mathematical concepts derived from Einstein's theory of general relativity. Over the past decades, growing astrophysical evidence has revealed the existence of black holes through multi-wavelength observations [\refcite{kr95_agn,kh13_agn}]. Active galactic nuclei (AGN), powered by accretion disks around black holes, emit radiation that can far exceed the total luminosity of the stars in their host galaxies, making them visible even at  the distant universe [\refcite{mortlock11_highz,fan23_agn}]. Furthermore, binary black holes can stir spacetime and generate gravitational waves that ripple across the cosmos, as first detected by the LIGO observatory [\refcite{LIGO16}]. A recent breakthrough in observational technology, namely the Event Horizon Telescope, has allowed astronomers to capture shadow images of supermassive black holes in galactic centers for the first time [\refcite{eht19_bh}]. These landmark discoveries represent major milestones in modern physics and have been recognized with several Nobel Prizes.

The theoretical formation of cosmic black holes is conceptually straightforward: if a massive amount of matter is compressed into an extremely small volume, its gravitational field becomes so strong that not even light can escape, leading to the formation of a black hole [\refcite{mtw}]. However, creating such conditions in the laboratory is practically impossible because of the dominant role of electromagnetic and strong nuclear forces within atoms. In contrast, gravity is the dominant force on astrophysical scales, enabling the collapse of massive stars or star-forming clouds to potentially produce black holes.

According to stellar evolution theory, massive stars of $> 10$ solar masses (\Msun) can undergo core collapse at the end of their lives, resulting in supernova explosions and possibly forming black holes with masses of $\sim 6$–$51~M_\odot$ and $> 130 \Msun$ [\refcite{brk67,hw02,wbh07,kasen11,chen14a,chen14c,woosley_rmp,zwh08_collapse,woos17,chen17a,chen17c}]. However, how these stellar-mass black holes grow into supermassive black holes (SMBHs) with millions or even billions of solar masses (\Msun) remains a major unsolved question in astrophysics [\refcite{lauer2007_smbhmasses,titans}].
For a deeper understanding of the background and evolution of astrophysical black holes, we recommend several comprehensive reviews [\refcite{kh13_agn,fg11,mf01_model,kohei20_smbh}]. 

In this short review, we focus specifically on the origins of SMBHs located at the centers of galaxies. In Section~2, we summarize observational evidence for the existence of SMBHs and discuss theoretical models of their formation. Section~3 explores the co-evolution of rapidly growing SMBHs and their host galaxies, highlighting key challenges in modeling their growth. Finally, in Section~4, we present a summary and offer future perspectives.

\section{The Origin of Supermassive Black Hole}

The existence of SMBHs in galaxies throughout cosmic history is well established [\refcite{kr95_agn,lauer2007_smbhmasses,gial15}]. However, the mechanisms underlying their formation remain one of the most fundamental unresolved questions in astrophysics. Understanding the origin and rapid growth of these black holes is essential for constructing a comprehensive picture of cosmic evolution.

The formation of SMBH seeds is believed to be closely related to the death of massive stars that formed in the early universe. According to modern cosmological models, the first generation of stars—so-called Population~III (Pop~III) stars—are predicted to have formed several hundred million years after the Big Bang. These stars were born within the gravitational potential wells of dark matter halos with typical masses of $\sim 10^6~\Msun$, which enabled the collapse of primordial gas composed of $\sim 76\%$ hydrogen and $\sim 24\%$ helium [\refcite{bl04c,ks23}]. In the absence of heavy elements (metals), molecular hydrogen served as the primary coolant. However, it was not very efficient, leading to higher gas temperatures and, consequently, a higher Jeans mass in Pop~III star-forming clouds compared to present-day star-forming regions. This environment favored the formation of extremely massive stars, with characteristic masses in the range of $100$–$1000~\Msun$ [\refcite{bl03,hir15}]. At the end of their lifespans, some of these very massive stars are expected to have collapsed directly into black holes, potentially seeding the formation of SMBHs.

To date, three primary scenarios have been proposed for the origin of SMBHs at galactic centers [\refcite{rees84,vol12}]:

\begin{enumerate}
    \item {\bf Primordial Stellar Black Hole Seeds with Extreme Accretion}\\
    Cosmological simulations suggest that the first generation of stars Population~III stars—had characteristic masses in the range of $100$–$1000~M_\odot$.These massive stars could end their lives as black holes with masses of several hundred solar masses. If such black hole seeds are embedded in gas-rich environments and can sustain the highest accretion rate, they may grow efficiently over cosmic time into the SMBHs observed in the centers of galaxies today.\\

    \item {\bf Supermassive Stellar Seeds with Moderate Accretion}\\ Within the first galaxies, dark-matter halos of $10^8-10^9\,\Msun$ may have facilitated the formation of Supermassive Stars (SMSs) with masses ranging from $10^4$ to $10^6~M_\odot$ [\refcite{wise12,jet13,lf16,hir17,chon17a,wise19}]. These SMSs could collapse directly into black holes of comparable mass. Given their large initial seed masses, such black holes require much less extreme growth conditions than stellar-mass seeds to reach the observed supermassive scales. \\

    \item {\bf Mergers of Pop~III Stellar Black Holes with Strong Accretion} \\
    The first galaxies likely contained numerous massive stars with initial masses of $30$–$100~M_\odot$, which would eventually collapse into black holes [\refcite{wise12,greif09,greif11,ss13}]. Through dynamical interactions and gravitational mergers in dense stellar environments, these black holes could coalesce into mass black holes of $\sim 10^3~M_\odot$. The resulting mass of seed black hole is between scenarios (1) and (2); it needs a strong gas accretion to grow into SMBHs.
\end{enumerate}

\begin{figure}[h]
\hspace*{-2.5cm}\includegraphics[width=1.4\columnwidth]{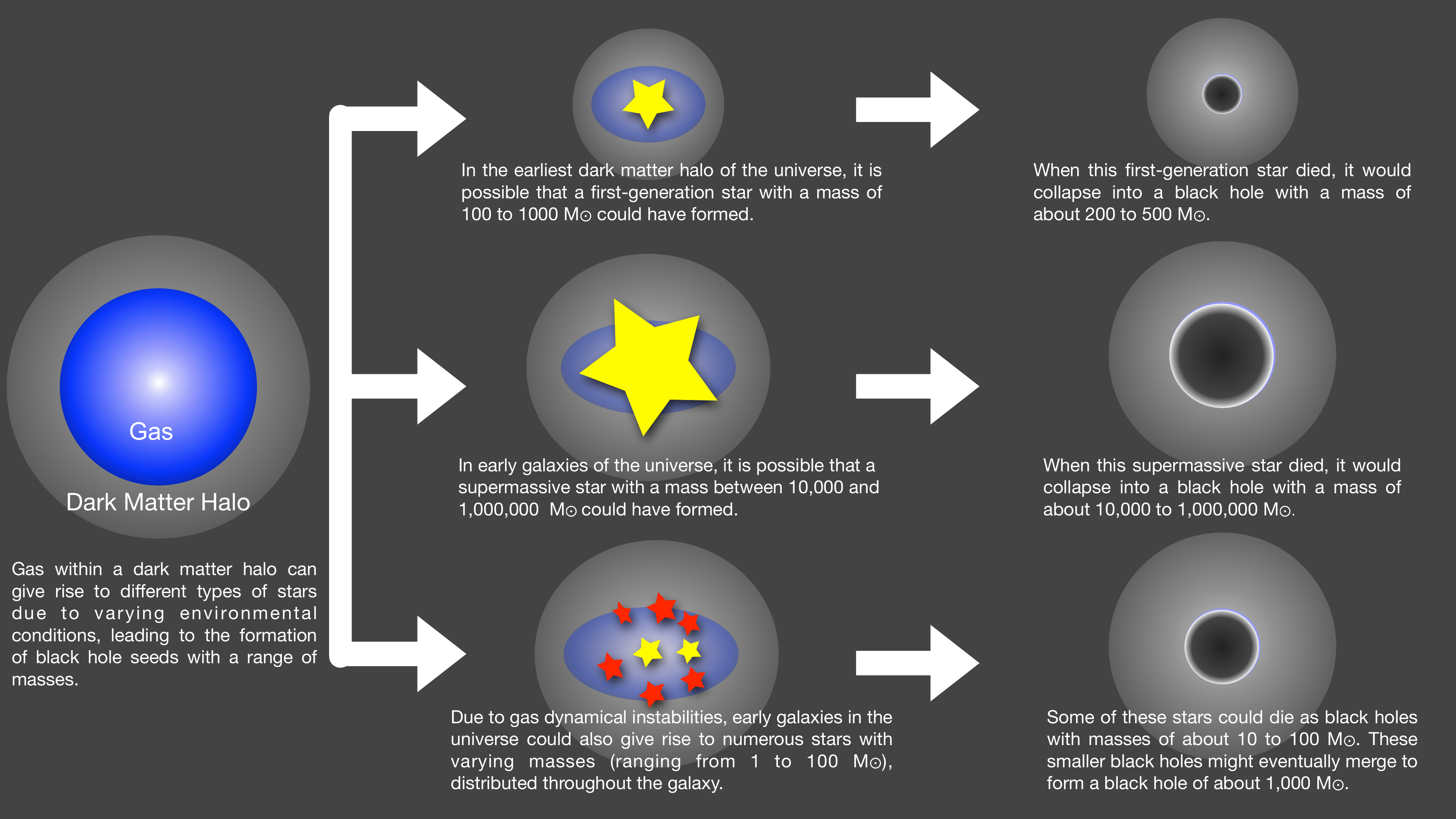} 
\caption{Three pathways for seeding the SMBHs \label{fig:pathway}.}
\end{figure}

We summarize the three main pathways for the formation of SMBH seeds in Figure~\ref{fig:pathway}. Among these, current mainstream research increasingly favors the second mechanism—supermassive stellar seeds—as the most plausible origin of SMBH seeds. In particular, the direct collapse of SMSs into black holes [\refcite{fowler64,fowler66,bet84}] has emerged as a promising scenario for seeding SMBHs [\refcite{begel08_smbhacc,begel10,titans}].

Regardless of the specific formation mechanism, sustained gas accretion is essential for black hole growth. Observational evidence indicates that SMBHs with masses of several hundred million solar masses were already in place within the first billion years after the Big Bang. This rapid growth of SMBHs is strongly associated with their host galaxies.

\section{Rapid Growth of SMBH and Their Coevolution with Host Galaxies}
\lSect{xx}

Even SMBH seeds can form with masses ranging from $10~\Msun$ to $10^6~\Msun$. These black hole seeds need to grow to masses of millions, or even billions, of solar masses, as suggested by the observation of AGNs at the time when the universe was less than one billion years old. The rapid growth of SMBHs is required to explain such observations and necessitates a continuous supply of gas, which is closely linked to the host galaxy. Understanding black hole growth thus requires a fundamental understanding of the mass accretion process.

\begin{figure}[h]
\includegraphics[width=1.\columnwidth]{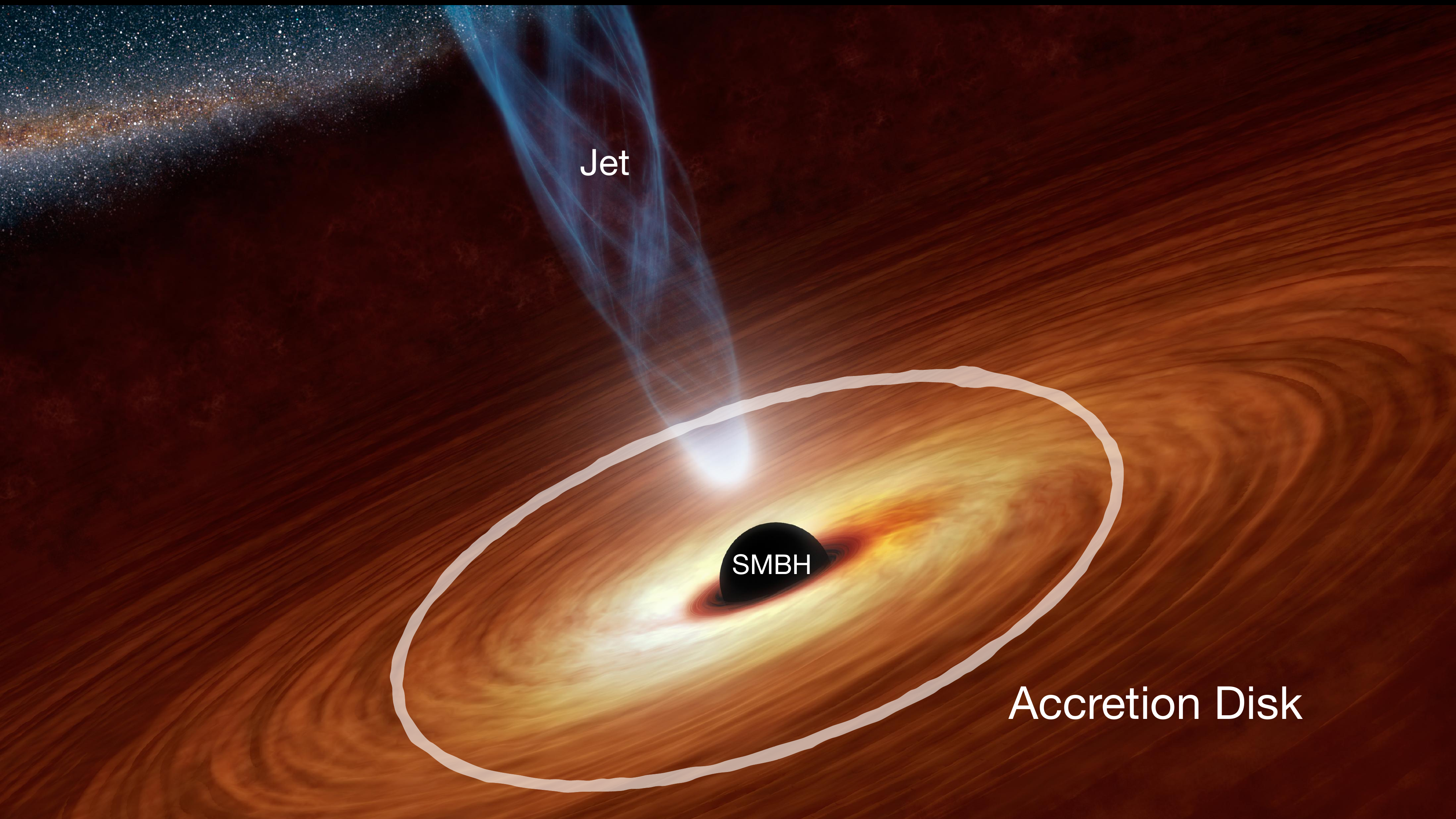} 
\caption{Schematic of the SMBH accretion disk and its jet. The accreting gas onto the SMBH first forms an accretion disk. Due to the strong gravitational field of the SMBH, a large amount of potential energy is released, heating up the accretion disk. This process leads to the emission of strong luminosity and winds. If the SMBH is spinning and interacts with the magnetic field from the disk or the surrounding medium, a jet can be launched. The disk wind, radiation, and jet represent feedback from the SMBH that can influence the evolution of its host galaxy and the surrounding intergalactic medium. (Background image credit: NASA/JPL-Caltech)
\label{fig:acc}}
\end{figure}

As the nature of dark matter remains unknown, studies of black hole accretion mainly focus on gas accretion. Gas surrounding black holes forms an accretion disk because of the gravitational influence of the black hole. Material in the inner regions of the accretion disk eventually crosses the event horizon, driving black hole growth. As matter falls into the disk, it releases a large amount of energy, $\sim 10\%$ of its mass multiplied by the square of the speed of light, that is 100 times more than energy released from the nuclear fusion in stars. Consequently, when SMBHs accrete rapidly, they emit more energy than the combined starlight of their host galaxies, becoming the primary source of galactic luminosity, as shown in Figure~\ref{fig:acc}.

Accretion disks release vast amounts of energy, and this energy feedback can prevent further gas accretion once it reaches a certain intensity. This feedback mechanism imposes a growth limit—the Eddington accretion rate. The \textbf{Eddington accretion rate} is the maximum steady rate at which matter can accrete onto a compact object, such that the outward radiation pressure from the emitted photons balances the inward gravitational force. It is derived from the \textbf{Eddington luminosity}, which is given by
\[
L_{\text{Edd}} = \frac{4 \pi G M c}{\kappa},
\]
where \( G \) is the gravitational constant, \( M \) is the mass of the accreting object, \( c \) is the speed of light, and \( \kappa \) is the opacity, typically taken to be the Thomson opacity for fully ionized hydrogen. The corresponding Eddington accretion rate is defined as
\[
\dot{M}_{\text{Edd}} = \frac{L_{\text{Edd}}}{\eta c^2},
\]
where \( \eta \) is the radiative efficiency, usually assumed to be around 0.1 for a standard thin accretion disk around a black hole. Substituting typical values yields an approximate expression:
\[
\dot{M}_{\text{Edd}} \approx 2.2 \times 10^{-8} \left( \frac{M}{M_\odot} \right) \left( \frac{0.1}{\eta} \right) M_\odot~\text{yr}^{-1}.
\]
This sets a fundamental limit for steady, radiatively efficient accretion: the Eddington accretion rate, $\dot{M}_{\text{Edd}}$, which is proportional to the mass of the accreting black hole. More massive black hole seeds can therefore grow faster, facilitating the formation of SMBHs. However, under certain conditions, the mass accretion rate of SMBHs may temporarily exceed the Eddington limit during specific duty cycles [\refcite{inay16,wl12,takeo2019_supereddington}].

\begin{figure}[h]
\includegraphics[width=1.\columnwidth]{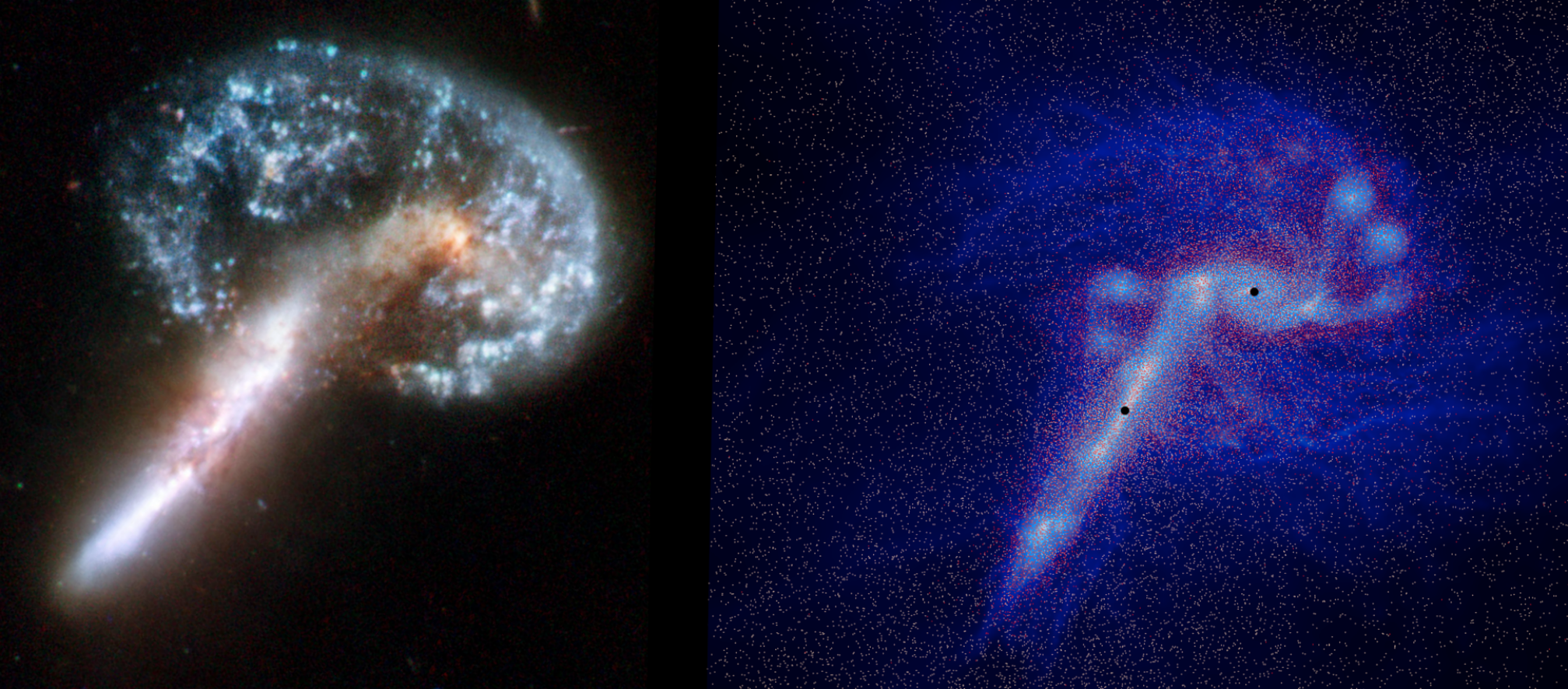} 
\caption{Left image: Arp 148 is a galaxy with a peculiar structure formed after the collision of two galaxies, where a large amount of material falls into the center of the galaxies, giving rise to its unique appearance (Image source: NASA, ESA, the Hubble Heritage Team (STScI/AURA)-ESA/Hubble Collaboration). Right image: a simulation of the formation of Arp 148. When two disk galaxies collide head-on, a significant amount of molecular clouds is accreted into the central region, providing fuel for the central black hole. Simultaneously, this process triggers a burst of star formation in the central region of the galaxy. These simulation results match well with the observed features of Arp 148 (Image source: Chi-Hong Lin\cite{lin22})}
\label{fig:bh_merge}
\end{figure}

Another challenge lies in providing sufficient gas environments to sustain black hole growth. Since SMBHs reside in the centers of galaxies, their growth must be closely tied to their host galaxy's gas supply. Hubble Space Telescope images reveal galaxies in various shapes, including disks, spirals, ellipticals, and irregular galaxies, such as the Mice Galaxy shown in Figure~\ref{fig:bh_merge}. These peculiar galaxies often result from galaxy collisions. Galaxy mergers are common in the universe and play a crucial role in the evolution of galaxies. During galaxy interactions, gravitational tidal forces disrupt the dynamical balance of stars and gas, driving some of the gas toward the center of the galaxy. This process triggers new star formation and supplies the central black hole with gas, promoting its growth. Such interactions are a key element in both galaxy evolution and the formation of SMBHs.

Pioneering studies of [\refcite{vhm03_merger,springel05_merger,volonteri2005_bhgrow}] used numerical simulations to investigate galaxy mergers and the evolution of their central SMBHs. Their results showed that some gas flows toward the galaxy center, stimulating black hole growth. However, they concluded that this growth efficiency alone was insufficient to explain the observations.Since then, numerous studies have explored black hole growth in the context of galaxy mergers [\refcite{di2005_merger,hopkins06_galaxy,king2015_mergerfeedback,tung24}]. Despite various methods to enhance black hole accretion, simulations still struggle to match observational data, highlighting the complexity of this phenomenon and the need for further research.

SMBHs accrete cosmic gas, which is primarily composed of hydrogen and helium. Depending on the physical conditions, this gas can exist in ionized, neutral, or molecular phases. These phases exhibit distinct behaviors, much like the different states of water (gaseous, liquid, and solid). The density and temperature between dense molecular clouds and low-density ionized gas can differ by several orders of magnitude. Particles of high-density molecular cloud experiences greater dynamical friction, a phenomenon first described by Chandrasekhar in 1943 [\refcite{chandra43}]. In this process, gravitational interactions with surrounding particles reduce momentum and kinetic energy, driving massive particles toward the system's center. This explains why SMBHs are usually located at the centers of galaxies. However, previous studies of galaxy mergers and black hole growth often neglected the effects of dynamical friction. Incorporating this effect, especially in the context of galaxy collisions, reveals that dense molecular gas gathers more efficiently at galaxy centers than neutral or ionized gas. This may explain why earlier simulations failed to replicate the rapid growth of black holes observed in some galaxies.

Recent work of [\refcite{lin22}] conducted a detailed study on rapid black hole growth using high-resolution galaxy collision simulations and examined black hole growth and its effects on host galaxies, considering the effect of dynamical friction from molecular clouds for the first time. [\refcite{lin22}] demonstrated that black hole growth primarily stems from the accretion of molecular clouds during galaxy mergers. During a galaxy merger, the equilibrium of gas and stars in both galaxies is disrupted, and dynamical friction drives dense molecular gas toward the galaxy center. This process not only triggers the burst of star formation but also provides the black hole with the necessary fuel for rapid growth, as shown in Figure \ref{fig:bh_merge}. This mechanism explains how black holes, with initial masses of several million solar masses, can grow to several hundred million solar masses within a billion years, matching observations of high-z AGNs.

\section{Key Issues on Modeling frontier of SMBH Seeds}

Although significant progress has been made in the past two decades, several substantial challenges persist in modeling SMBH formation and evolution. We highlight some of the key issues on modeling the seed formation as follows:

\begin{itemize}
    \item The absence of high-resolution cosmological simulations capable of capturing the specific conditions necessary for SMS formation.
    \item Existing stellar evolution models for SMSs remain incomplete, often neglecting key physical processes such as accretion physics, rotation, and binarity.
    \item Statistical predictions for the SMS mass function and the corresponding SMBH seed mass distribution are still underdeveloped.
\end{itemize}

A particularly compelling aspect of this problem is that SMSs with masses exceeding $10^4$ solar masses could have collapsed into SMBH seeds without leaving direct observational evidence. Therefore, it is crucial to explore the observational signatures associated with the collapse of SMSs. This includes investigating electromagnetic and gravitational wave signals expected from gas accretion during SMS collapse, general relativistic supernovae [\refcite{chen14b,moriya21}], and binary SMS mergers.

\section{Summary and Perspective}

Recent detections of Little Red Dots (LRDs) in the JWST era suggest the presence of highly obscured young AGNs at redshifts of $z > 5$ [\refcite{koce23_lrd,green24_lrd}]. LRDs provide compelling evidence for the rapid growth of nascent SMBHs. These SMBHs may originate directly from the seed black holes discussed in previous sections and serve as progenitors of SMBHs. However, explaining the observed high redshift AGNs of $z \sim 7 $ [\refcite{mortlock11_highz}] requires the presence of massive seed black holes that likely formed from SMSs at $z>10$.

Establishing the existence of SMSs in the early universe would thus provide crucial insights into the origins of SMBHs observed within the first billion years after the Big Bang. A comprehensive investigation of SMS physics, their formation environments, and their eventual collapse into black holes using state-of-the-art simulations will significantly enhance our understanding of SMBH evolution. Furthermore, predicting potential observational signatures associated with SMS collapse, such as exotic supernovae and merging binary black holes, could serve as indirect evidence of their existence. These observational signatures may be detectable through gravitational wave signals and electromagnetic emissions.

With the launch of the James Webb Space Telescope (JWST) and the upcoming next-generation 30-meter-class telescopes, unprecedented high-resolution galaxy observations will probe the rapid growth of SMBHs and their role in galaxy evolution. Future gravitational wave observatories, such as the Laser Interferometer Space Antenna (LISA), will detect gravitational wave signals from binary SMBH seeds in the early universe. Coupled with next-generation artificial intelligence (AI) techniques and supercomputer simulations, the synergy between observations and theoretical models has the potential to unravel the origins of SMBHs and their connection to galaxy evolution.

\section*{Acknowledgments}
This research is supported by the National Science and Technology Council, Taiwan, under grant No. MOST 110-2112-M-001-068-MY3, NSTC 113-2112-M-001-028-, and the Academia Sinica, Taiwan, under a career development award under grant No. AS-CDA-111-M04. The author acknowledges the support of the Humboldt Research Fellowship for Experienced Researchers from the Alexander von Humboldt Foundation. This research was supported in part by grant NSF PHY-2309135 to the Kavli Institute for Theoretical Physics (KITP) and grant NSF PHY-2210452 to the Aspen Center for Physics. Our computing resources were supported by the National Energy Research Scientific Computing Center (NERSC), a U.S. Department of Energy Office of Science User Facility operated under Contract No. DE-AC02-05CH11231 and the TIARA Cluster at the Academia Sinica Institute of Astronomy and Astrophysics (ASIAA).

\bibliographystyle{mpla}

\end{document}